\documentclass{emulateapj}
\usepackage{color}
\usepackage{natbib}
\usepackage[normalem]{ulem}  

\newcommand{\dens}{g~cm$^{-3}$}

\slugcomment{Not to appear in Nonlearned J., 45.}

\shorttitle{Surface and core detonations in rotating white dwarfs}

\shortauthors{Garc\'\i a-Senz, Cabez\'on \& Dom\'\i nguez}

\begin{document}


\title{Surface and core detonations in rotating white dwarfs}  


\author{D. Garc\'\i a-Senz\altaffilmark{1,2},   
R.M. Cabez\'on\altaffilmark{3}, I. Dom\'\i nguez\altaffilmark{4} 
}

\altaffiltext{1}{Departament de F\'\i sica, UPC, 
(EEBE) Eduard Maristany 10-14, 08019 Barcelona, Spain; domingo.garcia@upc.edu} 
\altaffiltext{2}{Institut d'Estudis Espacials de Catalunya, Gran Capit\`a 2-4, 
 08034 Barcelona, Spain}
\altaffiltext{3}{sciCORE, Universit\"at Basel. Klingelbergstrasse, 61, 4056 Basel, Switzerland; ruben.cabezon@unibas.ch}
\altaffiltext{4}{Departamento de F\'\i sica Te\' orica y del Cosmos, Universidad de Granada, E-18071 Granada, Spain}



\begin{abstract}

The feasibility of the Double Detonation mechanism, - a surface Helium-detonation followed by the complete carbon detonation of the core -, in a rotating white dwarf with a mass $\simeq 1 M_{\sun}$ is studied using three-dimensional hydrodynamic  simulations. Assuming rigid rotation, the rotational speed is taken high enough as to considerably distort the initial spherical geometry of the white dwarf. Unlike spherically symmetric models, we found that when helium ignition is located far from the spinning axis the detonation fronts converge asynchronically at the antipodes of the igniting point. Nevertheless, the detonation of the carbon core still remains as the most probably outcome. The detonation of the core gives rise to a  strong explosion, matching many of the basic observational constraints of Type Ia Supernova. We conclude that the Double Detonation mechanism also works when the white dwarf is spinning fast. This confirms the sub-Chandrasekhar-mass models and, maybe some Double Degenerate models (those having some helium fuel at the merging moment), as appealing channels to produce Type Ia Supernova events.            
 
\end{abstract}




\section{Introduction}

 A challenging task in astrophysics is to unveil the progenitors and explosion mechanisms of Type Ia supernovae (SNe Ia). Nowadays, observational and theoretical arguments point to two major production channels for these explosions, called the Single Degenerate (SD) \citep{whe73} and Double Degenerate (DD) \citep{ibe84} scenarios (for reviews  see e.g. \cite{hil13,mao14}. The precise fraction of SNe Ia coming from each channel is still a matter of a vivid debate. 
 
A particular class of the SD models which have recently deserved  attention are those known as the Double Detonation (DDet) of a white dwarf (WD) with a mass well below the Chandrasekhar-mass limit. In the DDet model a carbon-oxygen (CO) white dwarf with masses $\simeq 0.8 - 1.1$~M$_\sun$~incorporates helium through the accretion from a companion star. Under the appropriate conditions \citep{woo94}, the helium detonates above the edge of the CO core, which in turn induces a second detonation of carbon, thus producing a Type Ia supernova.

There was a time when these sub-Chandrasekhar-mass explosion models (hereafter, subCh-mass models) had some success, because they were able to reproduce many supernova observables, especially the explosion energy and gross nucleosynthetic production for sub-luminous events \citep{woo86, woo94, liv95, gar99}. At the same time the DDet explosion mechanism \citep{liv91} was better understood than the subsonic deflagration wich powers, at least initially, the explosion in the Chandrasekhar-mass models \citep{nom84, kho91, nie00}. Nevertheless, the subCh-mass explosion models suffer from several drawbacks. The more acute of them is that the synthetic spectra does not match observations because they predict too much high-velocity $^{56}\mathrm{Ni}$~in the external layers, which also produce blue colors at maximum light due to radioactive heating, in contrast with observations \citep{hof96, nug97}.

The situation changed when it was realized that the Double Detonation mechanism could be at work even in helium layers as  thin  as $\simeq 10^{-2}$~M$_{\sun}$~\citep{bil07} so the nickel problem vanishes. At the same time, it was realized that the observed SNe Ia rates and delayed time distributions could not be reproduced assuming only  SD and DD Chandrasekhar-mass explosions, while including SD and DD subCh-mass explosions may solve the problem  \citep{bad12, rui11, mao14}. Moreover, it has been recently claimed \citep{blo17,gol18} that the faint end of the Phillips relation \citep{phi93, phi99} could only be reproduced with subCh-mass explosions (but see also \cite{hof17}. 

Recent multidimensional simulations of the DDet scenario have been carried out by 
\cite{sim07,sim10,sim12,fin07,fin10} in 2D (igniting in a point makes the problem axisymmetric) as well as in 3D \citep{mol13}, in this last case to discern the outcome of multipoint ignitions. All of them concluded that the Double Detonation mechanism is robust, being able to successfully cope with a variety of helium-shell masses and symmetric and non-symmetric initial conditions. 

Despite the fact that accretion or merging scenarios imply, up to some degree, rotation of the exploding WD, the number of SD calculations that incorporate the effects of rotation in the explosion, is really scarce. Fast spinning white dwarfs with masses $1.46M_\sun \le M\textsubscript{WD} \le 2.02 M_\sun$~were considered by \cite{pfa10b,pfa10a}, who tried to explain the differences in the peak luminosity as a function of the rotation strength. They  concluded, however, that the match of the deflagration models with observations becomes worse for rotating WDs.  Conversely, if the star explodes following a detonation, Super-Chandrasekhar-mass models in fast rotation may explain some basic features of super-luminous Type Ia events. The impact of a moderate amount of rotation in the gravitational confined detonation (GCD) model \citep{ple04} has been explored by \cite{gar16}, who concluded that rotation is a necessary ingredient to discern if the CO core detonates or not. 

In this work we investigate, for first time, the feasibility of the DDet mechanism when a white dwarf with a mass $\simeq 1M_{\sun}$~is rotating rapidly. This is especially relevant in this case because the secondary detonation of the CO core requires the focusing of the shock waves produced during the He-shell detonation onto a small region at the symmetry axis. We investigate to what extent such wave convergence might be hampered in rotating models, especially when the helium ignition takes place in a point-like region far from the spinning axis. Additionally, our models predict several properties that could be compared with observations, like kinetic energies, nuclear yields and asymmetries produced by the explosion mechanism.    

In Section 2, we describe the main features of the spinning white dwarfs considered in this work. In Section 3, we comment on the main features of the hydrodynamics code (SPHYNX) used in this work, the initial setting and the method to build stable rotating white dwarfs in rigid rotation (which is described with more detail in the Appendix).  We give a detailed description of the hydrodynamic evolution and nucleosynthesis during the detonation of the helium shell in Section 4. The detonation of the core and its consequences are described in Section 5.  Finally, Section 6 summarizes the main conclusions of our work.

\section{Rotation of accreting white dwarfs}
\label{SecRotation}
Conservation of angular momentum makes compact objects prone to have large spinning velocities. In particular, for compact binary systems the rotational velocity of the accreting WD benefits from the transfer of angular momentum from the accretion disc, being even able to approach the centrifugal threshold \citep{yoo04A}. In the case of subCh-mass models of Type Ia supernova an upper limit of the rotation velocity can be inferred assuming that the angular momentum of the accreted shell is efficiently transferred to the underlying white dwarf. Thus, considering no angular momentum losses, a quantitative relationship between the amount of accreted matter and the normalized angular velocity, $\Omega=\omega_{acc}/\omega_{kepl}$~ of the WD can be built \citep{lan00},

\begin{equation}
\Omega=\frac{3}{4 r_g^2}\left[1-\left(\frac{M\textsubscript{WD,i}}{M\textsubscript{WD}}\right)^{\frac{4}{3}}\right]
\label{velrotWD}
\end{equation} 

\noindent
where  $M\textsubscript{WD,i}$~is the initial mass of the WD, prior to accretion, $M\textsubscript{WD}$~is the mass of the white dwarf,  $\omega_{acc}$~is the angular velocity gained from the accretion disc, $\omega_{kepl}$~is the keplerian angular velocity and $r_g$~is the gyration radius \citep{rit85},

\begin{equation}
r_g=0.452+0.0853\log\left(1-\frac{M\textsubscript{WD}}{M\textsubscript{CH}}\right);\quad M\textsubscript{WD}\le 0.95 M\textsubscript{CH}
\label{girationrad}
\end{equation} 

\noindent
where $M\textsubscript{CH}$~is the Chandrasekhar-mass limit. The Keplerian velocity is

\begin{equation}
\omega_{kepl}=\sqrt\frac{GM\textsubscript{WD}}{R^3\textsubscript{WD}}  
\label{wkepl}
\end{equation}

According to the published literature on subCh-mass  models, the thickness of the helium shell, $\Delta M_{He}$, at the moment of the explosion is within the range $0.01M_{\sun}\le \Delta M_{He} \le 0.15 M_{\sun}$~\citep{fin10,sim12,mol13}. Considering $M\textsubscript{WD}=1$~M$_{\sun}$~and $M\textsubscript{WD,i}=0.85$~M$_{\sun}$ in Eq.(\ref{velrotWD}), it results in $\Omega\simeq 0.9$. Such large value would bring the WD close to its centrifugal limit and, as a consequence, the initially spherical geometry will evolve into an oblate spheroid, which may have an impact in the outcome of the explosion.  A heuristic calculation may help to select the adequate candidates for the hydrodynamic simulations of surface detonations in rotating WDs. Firstly, we set the minimum density, $\rho_{He}$, able to support a steady Helium-detonation. According to previous studies $\rho_{He}\ge 10^6$~\dens~ \citep{woo94,mol13,hol13}. We  choose $\rho_{He}=1.6~10^6$~g~cm$^{-3}$~as the nominal density at the core-envelope interface at the moment of the explosion, following \cite{mol13}.  We then integrate the structure equations of a WD for a grid of central densities in the range $10^7\le\rho_c\le 4~10^8 $~\dens~at constant temperature $10^6$~K, and we switch the chemical composition from $X_C=X_O=0.5$ to $X_{He}=1$~when $\rho\le\rho_{He}$. Such switch marks the edge between the CO core and the He-envelope. In this rough approach,  $M\textsubscript{WD}$, $\Delta M_{He}$, $\Omega$ and $\omega_{kepl}$,  depend exclusively on the adopted central density at the moment of explosion. 

\begin{figure*}
\includegraphics[angle=-90,width=1.0\textwidth]{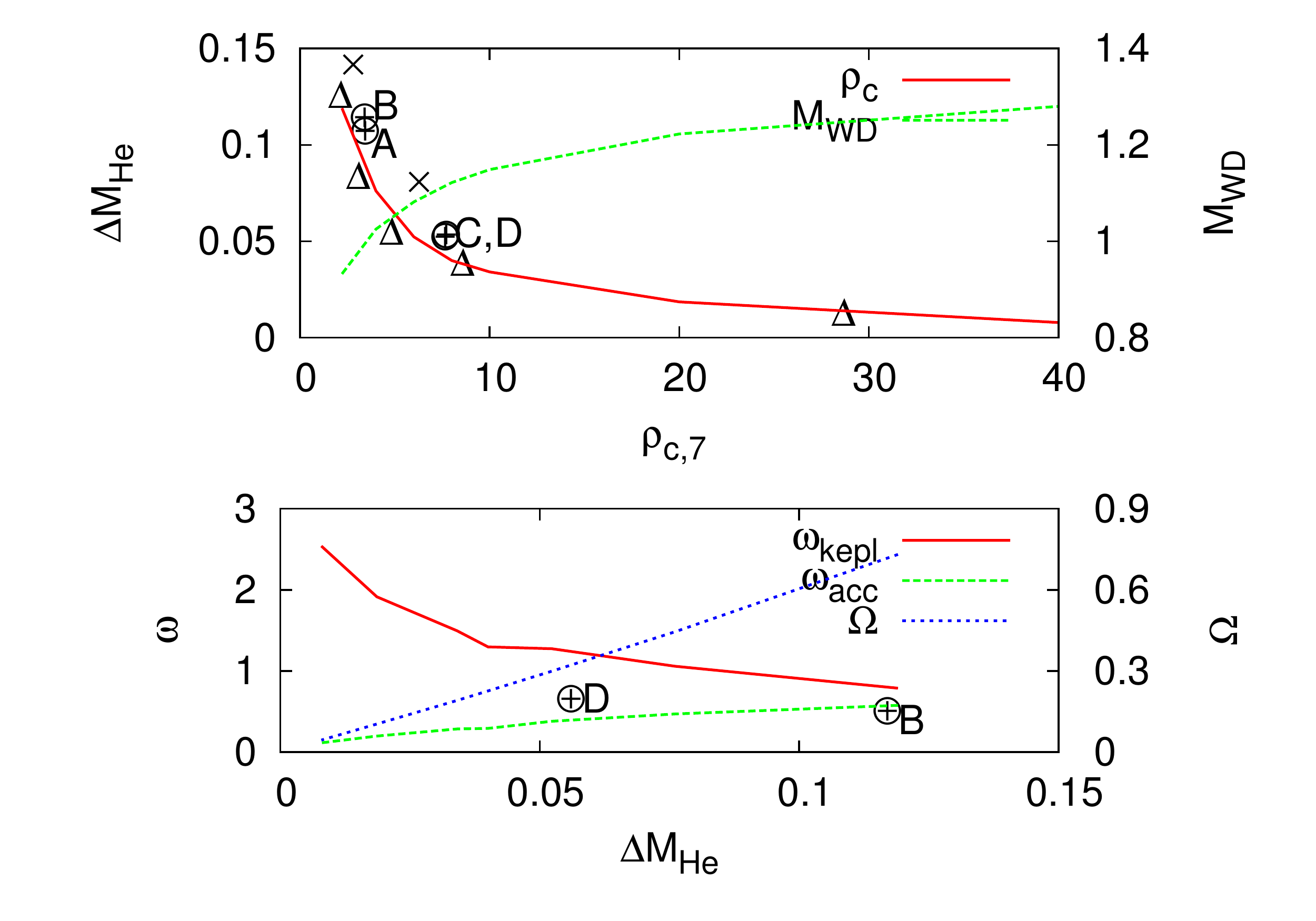}
\caption{Upper panel: mass of the helium shell ($\Delta M_{He}$, in $M_\sun$) on top of a CO core as a function of the central density for spherically symmetric models. Lower panel: rotational angular velocity of the WD ($s^{-1}$) as a function of the mass of the Helium-shell \ envelope. Symbols  $\triangle$, $\times$,~and $\oplus$, refer to explosion models reported in \cite{fin10}, \cite{mol13}, and Table~\ref{table1} in this work (models A, B, C and D), respectively.}  
\label{fig1}
\end{figure*}

The ensuing grid of models is depicted in Fig.~\ref{fig1}, where the upper panel gives the profile of $\Delta M_{He}$~and $M\textsubscript{WD}$~as a function of the central density, while the lower panel presents information concerning the angular velocity.  As it can be seen, the profile of $\Delta M_{He}$ is not longer linear. Tiny He-envelopes ($\simeq 0.02 M_{\sun}$) would require rather massive WD cores ($\simeq 1.2 M_{\sun}$) or, equivalently, large WD masses prior to accretion. On contrary, thick He layers ($\simeq 0.10 M_{\sun}$) would require a less massive WD ($\simeq 0.8 M_{\sun}$) prior to accretion. Such profile follows approximately the ($\rho_c, \Delta M_{He}$) relationship inferred from the data by \cite{fin10} and \cite{mol13} (triangles and crosses in Fig.~\ref{fig1} respectively).

The lower panel in Fig.~\ref{fig1} depicts the angular velocity, $\omega_{acc}$, of the WD after accreting $\Delta M_{He}$, the keplerian velocity $\omega_{kepl}$, as well as their ratio $\Omega=\omega_{acc}/\omega_{kepl}$. Any physically sound value of $\omega$~has to fulfill $\omega\le\omega_{acc}\le\omega_{kepl}$~where the equality $\omega=\omega_{acc}$~stands for conservative angular momentum transfer from the disc to the white dwarf. The symbols $\oplus$~in the figure indicate the location of models $A$, $B$, $C$ and $D$~described in Table~\ref{table1}. Models A and C are non-rotating models, while B and D are the corresponding rotating versions. We see that the angular velocity of model $B$ is in the  desired region of the diagram, albeit close to $\omega_{acc}$.  It is also worth noting that although rotation of model $D$~is neatly sub-keplerian its angular velocity is slightly above the $w_{acc}$~line. 

 White dwarfs are very compact and chemically homogeneous objects, so the transport of angular momentum is expected to be very efficient \citep{mae00, pir08, sai04} and the accreting WD may be treated as a rigid rotator. The presence of magnetic fields favors rigid rotation \citep{neu17}, although for non-magnetic sub-Chandrasekhar masses the final rotational state is not so well constrained \citep{gho17} and differential rotating WDs may end as Helium novae \citep{yoo04B}. We decide to adopt a practical approach and assume rigid rotation in all our models.  Models $B$~and $D$~therefore represent extreme cases in the sense that if the Double Detonation mechanism  works for them it will also work for any rotating model located below the $w_{acc}$~line in Fig.~\ref{fig1}. Additionally, the minimum observed period of a WD in a cataclismic variable is $P = 27.8$~s for WZ Sag \citep{pat80}. That period is larger than the value $P\simeq 5$~s obtained with Eqs. (\ref{velrotWD}),(\ref{girationrad}) and (\ref{wkepl}), with $M\textsubscript{WD}^i=0.85M_{\sun}, M\textsubscript{WD}=1M_{\sun}$~and $R_{WD}= 5000$~km, suggesting a non-conservative evolution during the accretion. Probably some fraction of the incoming angular momentum is lost during the recurrent, Nova-like, phenomena  associated to the surface flashes which transform the accreted hydrogen into helium. Also, the  polarization spectra of common (normal-Branch) SNe Ia explosions does not favor  large departures from the spherical geometry \citep{wan08}. All this suggests that the rigid body angular velocity obtained using Eqs. (\ref{velrotWD}) and (\ref{wkepl}) has to be taken as an upper limit. In this regard, we note that high, near-keplerian rotational velocities, may be achieved during the merging process of two white dwarfs in the DD scenario \citep{lor09,dan15}.

Additionally, the rotational velocities considered in models B and D in Table \ref{table1} are high enough as to leave some imprint in the geometry and the distribution of mass within the WDs. In that case, if the explosion mechanism and the main observables of the explosion do not appreciably differ from the spherically symmetric case, we can safely infer that rotation does not represent a problem for the viability of SNe Ia subCh-mass models.

\section{Hydrodynamic method and initial setting} 

Surface He-detonations on top of  massive rotating CO cores ($\ge 0.8 M_{sun}$) are an intrinsic 3D phenomena. During the explosion, the former helium shell is ejected with velocities $\ge 2~10^4$~km~s$^{-1}$,~so that the characteristic size of the object changes from the initial $R\simeq 5~10^3$~km to $\simeq 10^5$~km in few seconds. Such a large change in size, along with the multidimensional nature of the explosion, make Lagrangian methods, such as SPH, ideally suited to simulate these systems. Moreover, the addition of rotation renders this problem difficult to be studied using Eulerian hydrodynamics. To carry out the simulations we made use of the Integral-SPH (ISPH) hydrocode SPHYNX \citep{cab17}, conveniently adapted to handle explosive scenarios involving degenerated matter \citep{gar16}. SPHYNX is an state-of-the-art hydrocode with an improved algorithm to estimate gradients, which relies on an integral approach \citep{gar12} to the derivatives. It also makes use of the $sinc$~family of kernels \citep{cab08}, which are resistant to particle clustering, therefore allowing to increase the number of interpolating particles in the SPH summations to reduce the numerical noise.   

The physical processes included are very similar to those recently used by \cite{gar16} to study the GCD explosion mechanism. An efficient nuclear network evaluates the energy input and composition change due to nuclear reactions via an $\alpha-$chain, completed with carbon and oxygen binary reactions. The evolution of the species is calculated implicitly and coupled with the temperature, to ensure a smooth transition to the nuclear-statistical equilibrium (NSE) regime \citep{cab04}. Electron captures on protons and nuclei have been neglected because their impact  on the dynamics of the explosion is secondary. Note that central densities are more than two order of magnitude lower than explosion ignition densities in Chandrasekhar mass WDs.  Our EOS has the contributions of electrons \citep{bli96}, ions (including Coulomb and polarization corrections) and radiation.

All calculations reported in this paper assume that the thermonuclear ignition of the WD starts in a single spherical region located in the Helium-rich region, close to the core-envelope edge\footnote{\cite{mol13} also explored the impact of starting the He-detonation at some altitude above the interface, when the density is $\rho_{He}\simeq 1.6~10^6$~\dens}. Ideally, the size of such initial detonator is dictated  by the environmental physical conditions set during the pre-ignition state, especially by density and temperature peak values and profiles. However, current three-dimensional calculations do not have sufficient resolution to allow a self-consistent initiation of the explosion and, therefore, the Helium-detonation must be artificially triggered.

\begin{deluxetable*}{ccccrcccccccc}
\tablecaption{Main features of the initial models.}
\tablehead{
\colhead{Model}&\colhead{N}&\colhead{Ignition Altitude}& \colhead{$\omega_x$}&\colhead{Ign.axis}&\colhead{$\rho_c$}& \colhead{$\rho_{He}$}&\colhead{$h_c$}&\colhead{$h_{He}$}&\colhead{$M\textsubscript{WD}$}&\colhead{$\Delta M_{He}$}&\colhead{Oblateness} \cr
 &$10^6$~part & \colhead{km}&\colhead{$s^{-1}$}&\colhead{-}&\colhead{$10^7$\dens}&\colhead{$10^7$\dens}&\colhead{km}&\colhead{km}& \colhead{M$_{\sun}$}&\colhead{M$\sun$}&\colhead{f} \\
}
\startdata
A$_1$&$2.0$&$4200$ &$0.00$&$X$&$2.60$&$0.15$&$48$&$129$&$0.9590$&$0.1068$& $0.00$ \\
B$_1$&$2.0$&$4300$ &$0.50$&$X$&$2.57$&$0.11$&$49$&$142$&$1.0815$&$0.1140$& $0.35$ \\
B$_2$&$2.0$&$4550$ &$0.50$&$XY$&$2.57$&$0.11$&$49$&$142$&$1.0815$&$0.1140$&$0.35$ \\
B$_3$&$2.0$&$5000$ &$0.50$&$Y$&$2.57$&$0.11$&$49$&$142$&$1.0815$ &$0.1140$&$0.35$\\
B$_4$&$2.0$&$3900$ &$0.50$&$X$&$2.57$&$0.15$&$49$&$129$&$1.0815$ &$0.1533$&$0.35$\\
B$_5$&$2.0$&$4200$ &$0.50$&$XY$&$2.57$&$0.15$&$49$&$129$&$1.0815$ &$0.1533$&$0.35$\\
B$_6$&$2.0$&$4600$ &$0.50$&$Y$&$2.57$&$0.15$&$49$&$129$&$1.0815$ &$0.1533$&$0.35$\\
C$_1$&$4.0$&$3880$ &$0.00$&$X$&$6.82$&$0.15$&$ 29$&$102$&$1.1052$&$0.0520$&$0.00$ \\
D$_1$&$4.0$&$3700$ &$0.65$&$X$&$6.87$&$0.12$&$30$&$113$&$1.1872$ &$0.0532$&$0.21$\\
D$_2$&$4.0$&$4230$ &$0.65$&$Y$&$6.87$&$0.12$&$30$&$113$&$1.1872$ &$0.0532$&$0.21$\\
\enddata
\tablecomments{Columns show: model name, number of particles, initial bubble ignition altitude with respeect the center of the WD, angular velocity, location of the initial bubble (XY refers to a ignition at 45$^\circ$ in the X-Y plane), central density of the WD, density at the He-core interface, smallest smoothing length (i.e. highest spatial resolution) at the core and at the He layer, total mass of the WD (CO core + He envelope), mass of the He envelope, and oblateness factor as $f=\frac{a-b}{a}$, where $a\simeq 8000$~km and $b\simeq 5200$~km are the equatorial and polar radius in B-models. The radius of the igniting ball at the edge of the core is $R_b=250$~km in all models.   
}
\label{table1}
\end{deluxetable*}

\subsection{Implementation of rotation}

An accurate method to build rotating WDs in hydrostatic equilibrium within the SPH framework does not exist. We have developed and checked a relaxation procedure which is able to produce self-gravitational rotating white dwarfs in equilibrium. This topic is, by itself, of sufficient interest for the SPH community as to deserve a careful description and analysis, which is deferred to an upcoming publication. Nevertheless, the foundations of the method are described in the Appendix where we provide  the reader with some details on how we built the stable, rigidly rotating, white dwarfs considered in this work.    

\section{Hydrodynamic simulations}

\subsection{He-shell detonation: Evolution of the reference model.}

Our control model is A$_1$, in Table \ref{table1}. This is a non-rotating spherically-symmetric model of a WD with $M\textsubscript{WD}=0.9590~M_{\sun}$ and $\Delta M_{He}=0.107~M_{\sun}$. On the other hand, our reference models (B$_1$, B$_2$~and B$_3$) for rotating white dwarfs have a total mass $M\textsubscript{WD}=1.0815~M_{\sun}$. The  helium shell amounts $\Delta M_{He}=0.114~M_{\sun}$, similar to that of the spherically symmetric model A$_1$. In B-models, the WD is rotating as a rigid body around the X-axis, with a value $\omega_x=0.5$~s$^{-1}$. As quoted before, we have decided to explore an upper limit in terms of rotational velocity. We note that a non-magnetic massive WD with this high angular velocity has been observed \citep{mer15, pop18} in a binary system, although the origin of such rotation is still unclear. The unique difference among B-models is the location where the He-detonation starts: either aligned (B$_1$), at $45^\circ$ (B$_2$) or at $90^\circ$~(B$_3$) with respect the rotation axis. The outcomes of these calculations are compared to the control model A$_1$, with similar central density and mass of the helium shell.   

In model A$_1$, the detonation of helium is induced at the edge of the CO core, at a radius $r=4200$~km. Being three-dimensional calculations, models A$_1$, B$_1$, B$_2$, and B$_3$ have a relatively low resolution (see columns 8 and 9 in Table~\ref{table1}). Therefore, to build a sustainable detonation we artificially  incinerate all the helium fuel inside a sphere with radius $250$~km. After a while, a steady detonation wave emerges which rapidly incinerates the whole envelope of the white dwarf. The properties and evolution of the He-detonation have been investigated in numerous works in two and three dimensions \citep{liv95,gar99,sim10,mol13}. On the whole, all of them agree in that the most critical issue is the convergence of the surface detonations at the antipodes of the initial incinerated region. Such  convergence is so strong as to induce the detonation of the carbon layer at or below the convergence point. Finally, the detonation of the carbon propagates through the core and volatilizes the star (see Section \ref{coredet}). 

\begin{figure}
\includegraphics[angle=0,width=\columnwidth]{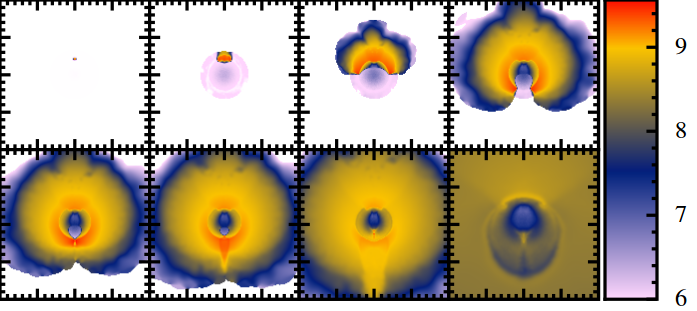}
\caption{Colormap of temperature in a XY slice, showing the explosion of the helium envelope of model $A_1$~in Table 1 at times $t=0.003, 0.200, 0.607, 1.000, 1.138, 1.255, 1.481$~and $3.647$~s, respectively. The collision of the detonation waves at the antipodes takes place between the fifth and sixth snapshots. The box size is $[-2:2]\times[-2:2]~10^4$~km The X and Y axes go from $-2~10^4$~km to $2~10^4$~km.}  
\label{fig2}
\end{figure}

\begin{figure}
\includegraphics[angle=0,width=\columnwidth]{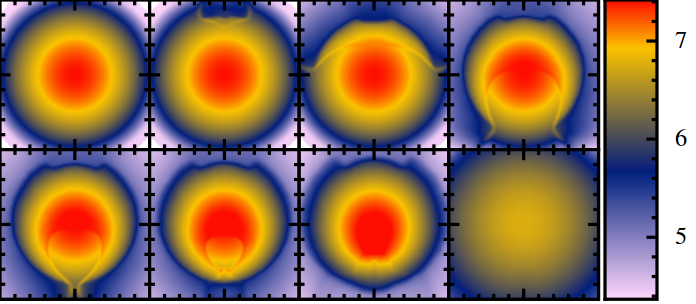}
\caption{Same as Figure \ref{fig2} but for density and zoomed in the central core of the WD. The box size is $[-5:5]\times[-5:5]~10^3$~km. }
\label{fig3}
\end{figure}

The evolution of model A$_1$ is in agreement with the findings of previous works. Actually, our results are similar to those of model A by \cite{mol13}. In Figs.~\ref{fig2} and \ref{fig3} we show the temperature and density colormaps at different times. In this calculation, the combustion of the carbon underneath was turned-off to maximize the density achieved during the collision at the antipodes. The convergence of the ashes of the He-detonation takes place at $t\simeq 1.18$~s, at an altitude of $r\simeq 4000$~km.  The collision of the ashes raises the temperature and density of carbon to $T\simeq 5.06~10^9$~K and $\rho\simeq 7.9~10^6$~\dens, more than enough to initiate the detonation of carbon, if nuclear reactions were switched-on \citep{sei09}. In Fig.~\ref{fig4}, we present the history of the maximum temperature achieved by any particle made of carbon and oxygen. The same figure also shows the corresponding density of that particle.  As we can see, there is a pronounced plateau in $T_{max}$~between times $1.16\leq t\leq 1.64$~s, where $T_{max}\geq 4~10^9$~K, and the CO mix is prone to detonate. Within this interval there is a prominent peak in density at $t\simeq 1.465$~s, where the  chances for carbon detonation are maximized. Such high values of density and temperature come after the convergence of the different shock waves at the symmetry axis, at an altitude $r\simeq 1650$~km.

The fate of the rotating models may rely on the precise location where the He-detonation starts. If the helium detonates just at the rotational axis (model $B_1$~in Table~\ref{table1}) a preferred symmetry line remains, joining the initial igniting spot with the center of the WD, and the evolution should not be very different to that of a spherically symmetric model (i.e. non-rotating). We note, however, that enforcing a similar $\rho_c$~and $\Delta M_{He}$ in rotating and non-rotating models produces slightly different ignition densities of helium at the core edge. As a result, the densities and temperatures in the converging region are higher in the spherically symmetric non-rotating model (Fig.~\ref{fig4} and Table~\ref{table2}).

\begin{deluxetable}{ccccrccccc}
\tablecaption{Main features during the detonation of the He-shell.}
\tablehead{
\colhead{Model}&\colhead{T$_{max}$}&\colhead{$\rho (T_{max})$}&\colhead{$E_{nuc}$}&\colhead{$^{44}\mathrm{Ti}$}& \colhead{$^{56}\mathrm{Ni}$} \cr
  &\colhead{$10^9$~K}&\colhead{$10^7$\dens}&\colhead{$10^{50}$~ergs}&\colhead{M$_{\sun}$}&\colhead{M$_{\sun}$} \\
}
\startdata
A$_1$&$5.06$&$0.79$&$1.83$&$2.24~10^{-2}$&$1.25~10^{-3}$ \\
B$_1$&$3.46$&$0.74$&$1.65$&$3.81~10^{-2}$&$1.18~10^{-4}$ \\
B$_2$&$3.79$&$0.66$ &$1.66$&$3.79~10^{-2}$&$1.48~10^{-4}$ \\
B$_3$&$4.13$&$0.70$ &$1.66$&$3.75~10^{-2}$&$1.90~10^{-4}$ \\
B$_4$&$4.86$&$0.92$ &$2.55$&$3.42~10^{-2}$&$1.24~10^{-3}$ \\
B$_5$&$4.50$&$0.95$ &$2.56$&$3.41~10^{-2}$&$1.35~10^{-3}$ \\
B$_6$&$3.63$&$0.82$ &$2.57$&$3.39~10^{-2}$&$1.63~10^{-3}$ \\
C$_1$&$4.54$&$1.36$ &$0.88$&$1.33~10^{-2}$&$2.45~10^{-4}$ \\
D$_1$&$3.60$&$0.92$&$0.80$&$1.82~10^{-2}$&$4.67~10^{-5}$\\
D$_2$&$3.70$&$0.78$&$0.81$&$1.80~10^{-2}$&$8.50~10^{-5}$ \\
\enddata
\tablecomments{Columns are: model name, values of $T_{max}$, $\rho (T_{max})$, total released nuclear energy, and titanium and nickel abundances exclusively coming from the detonation of the helium shell. The combustion of any particle belonging to the CO core has been artificially suppressed.}
\label{table2}
\end{deluxetable}

The values of $T_{max}$~and $\rho (T_{max})$ in the carbon  region for rotating B-models are shown in Fig.~\ref{fig4} (green, blue, and pink lines) and Table~\ref{table2}. As we can see, the profiles of temperature and density follow a trend similar to model A$_1$. Nevertheless, the temperature and density peaks in model B$_1$ are less pronounced. They are also delayed approximately $\Delta t\simeq 0.2$~s with respect to model A$_1$. According to the standard detonation criteria \citep{nie97,sei09} Carbon may detonate in model B$_1$ when $1.40\le t\le 1.5$~s. 

The evolution of models B$_2$ and B$_3$, igniting in an oblique line to the spinning axis, is a bit different. Several snapshots of the explosion of the He-layer of model B$_3$ (igniting at the equatorial plane) are depicted in Figures \ref{fig5}, \ref{fig6} and \ref{fig7}. The upper row of panels in Fig.~\ref{fig5} shows the temperature colormap in a XY-slice containing both, the rotational axis and the ignition point (a polar plane). Such polar plane is rotating with $\omega_x=0.5$~s$^{-1}$, so that it is a comoving projection plane. On the other hand, the lower row in the same figure shows the temperature in the equatorial plane as viewed from a non-rotating frame of reference. On the whole, the geometry of the oblated spheroid  desynchronizes the convergence of the ashes at the antipodes. This is more evident in the colormap of density, Fig.~\ref{fig6} and, especially in the close-up of Fig.~\ref{fig7}, which focuses around the convergence region. As we can see, the convergence is attained earlier in the polar plane than in the equatorial plane. Such shift in the converging times is purely geometrical, because in an oblated spheroid the polar geodesic has a length $l_{pol}= 2\pi a$, whereas the equatorial geodesic amounts $l_{eq}= 2\pi\sqrt{0.5(a^2+b^2)}$, where $a$ and $b$ are the equatorial and the polar radius. According to the values in Table~\ref{table1}, $l_{eq}/l_{pol}\simeq 1.18$;  any other geodesic has $l_g$ with  $l_{pol}\leq l_g\leq l_{eq}$. Admitting an isotropic distribution of detonation velocities, there is a continuous shift in the arrival times of the converging waves. Therefore, the strong focusing which characterizes models A$_1$ and B$_1$ is somehow lost in models B$_2$ and, more evidently, B$_3$. Still, the wave convergence at the antipodes is strong enough as to induce the detonation of carbon. Just imagine the picture from a rotating reference frame: as the detonation is supersonic the forces acting on a fluid element which goes through the shock front are much higher than the non-inertial forces, centripetal and Coriolis, which do not appreciably affect the propagation of the detonation wave \footnote{Nevertheless the inertial forces have some impact in the large-scale geometry of the explosions. In particular, the centrifugal barrier set by the rotation favors the elongated morphologies along the rotational axis \citep{pfa10b,pfa10a}}.  

The principal impact of rotation is to desynchronize the wave trains arriving to the antipodes of the igniting region. Such asynchronous wave arrival, however, does not necessarily reduce the peak temperature deep down the antipodes. According to Fig.~\ref{fig4} and Table~\ref{table2}, the largest value of $T_{max}$~for rotating models is actually achieved in model B$_3$ ($T_{max}=4.13~10^9$~K), followed by B$_2$ ($3.79~10^9$~K) and B$_1$~($3.46~10^9$~K) with densities $\rho\simeq 0.7~10^7$\dens~ in all three cases. If the $^{12}C+^{12}C$ reaction would have been switched-on, these temperatures and densities were high enough \citep{nie97,sei09} as to provoke the detonation of the core of the white dwarf (see Sect.~\ref{coredet}).    
We conclude that the ignition and detonation of carbon is the most probable outcome in all rotating models that we calculated. Therefore, the DDet mechanism appears to be robust: it not only works if helium is ignited in one or several points \citep{gar99,mol13} but also when the WD is rapidly rotating.  
              
The yields produced during the detonation of the helium shell are shown in Table \ref{table3}. These yields are only approximate owing to the small size of the $14$-nuclei network used to track the He-detonation. The main limitation comes, however, from the low resolution achieved in the helium envelope which results in a large fraction of unburnt helium after the freezing of the reactions at $t\geq 2$~s. Compared to the spherically symmetric model A$_1$, the final abundance of $^{56}\mathrm{Ni}$ is approximately an order of magnitude lower in B-models. The higher production of nickel in the non-rotating model is due to: a) the slightly higher ignition density of helium in model A$_1$, b) the higher densities and temperatures achieved at the converging region in model A$_1$ (Fig.~\ref{fig4}) and c) fast rotators have a larger amount of mass 'stored' at low densities, which disfavors the production of IGE. In all cases, but especially in the rotating models, the more abundant ejected species are the radioactive $^{44}\mathrm{Ti}$~and $^{48}\mathrm{Cr}$. We note that the presence of Ti absorption lines in the near maximum spectra has been suggested as an indicator of the He-detonation-triggered scenario \citep{jia17}.

As pointed out in previous works by other authors \citep{sim12} the detonation of the He-shell alone would produce a sub-luminous event ($M_{bol}\simeq -16.5$) with a peculiar light curve dominated by the disintegration of $^{52}\mathrm{Fe}$ rather than $^{56}\mathrm{Ni}$ at early times.         

\subsection{Geometry of the ejected shell}
\label{geometryHe}

A point-like, edge-lit ignition of the helium envelope, followed or not by the complete detonation of the CO core of the WD, leads to a loss of the spherical symmetry which may be detected in polarization studies \citep{fin10,bul16,bul17}. We want to investigate if such loss of spherical symmetry is more pronounced in rotating WDs. In Fig.~\ref{fig8} we show the combined column density of radioactive $^{48}\mathrm{Cr}$+    $^{52}\mathrm{Fe}$+$^{56}\mathrm{Ni}$ for different models, at $t\simeq 8.3$~s, when the expansion is homologous. Such column density \footnote{Obtained and drawn with the public program SPLASH written by D. Price \citep{pri07}} is estimated assuming an artificial photosphere with local thickness $2\bar h$ (being $\bar h$ the average of the smoothing length) and projected onto three orthogonal observer planes YZ, XY, XZ respectively (being the plane YZ parallel to the equator of the WD). Because these radioactive elements are expanding homologously, their relative spatial distribution will not change afterwards. with time after $t\simeq 8.5$~s, up to the moment at which these elements begin to disintegrate several days after. 

The upper row panels in Figure \ref{fig8} depict the 'brightness' of the photosphere for the non-rotating model A$_1$. The distribution of radioactive $^{48}\mathrm{Cr}$+$^{52}\mathrm{Fe}$+$^{56}\mathrm{Ni}$~is not totally spherical when viewed perpendicularly to the polar direction (central and rightmost snapshots), with a larger concentration in the northern hemisphere. On another note, the distribution is rather smooth, free from pockets of $^{56}\mathrm{Ni}$ which characterize pure deflagration models \citep{gar05}. The impact of such asymmetric distribution of IGE and IME in the polarization of the spectra in subCh-mass models has been recently analyzed by \cite{bul16}. They conclude that the asymmetries are not large enough to produce significant levels of polarization ($\geq 0.5\%$) in the spectra. We note that the polar view (leftmost snapshot) is totally symmetric, as expected. 

Figure~\ref{fig8} also shows the column density of the radioactive elements synthesized during the He-detonation of rotating models B$_1$, B$_2$, and B$_3$. In particular, models B$_1$ and B$_2$ look similar to the control model A$_1$, but they are slightly more elongated in the direction of the rotational axis (central and rightmost columns in Fig.~\ref{fig8}). Such anisotropic distribution of the burning products is due to the angular momentum barrier set by the rotation, which is stronger in the equatorial direction \citep{pfa10a}. Interestingly, the distribution of radioactive elements in model B$_3$ seems to be more spherical than in models B$_1$ and B$_2$, in those planes. When viewed from the polar axis (leftmost column in Fig.\ref{fig8}), models B$_1$ and B$_2$ look similar to A$_1$, but B$_3$ has a clear loss of spherical symmetry. Although the loss of spherical symmetry is larger than in the non rotating model, providing quantitative numbers for its impact on the polarization of the spectra is out of the scope of the present work.         

To sum up, the single detonation of the helium shell in a rotating $\simeq 1$~M$_\sun$ white dwarf would produce a sub-luminous event powered by the disintegration of $^{48}\mathrm{Cr}$+$^{52}\mathrm{Fe}$, and $^{56}\mathrm {Ni}$. The asymmetries in the distribution of nuclear species are larger than in spherically symmetric models, which probably will increase the level of polarization in the light curve and spectra.

\begin{figure}
\includegraphics[angle=-90,width=\columnwidth]{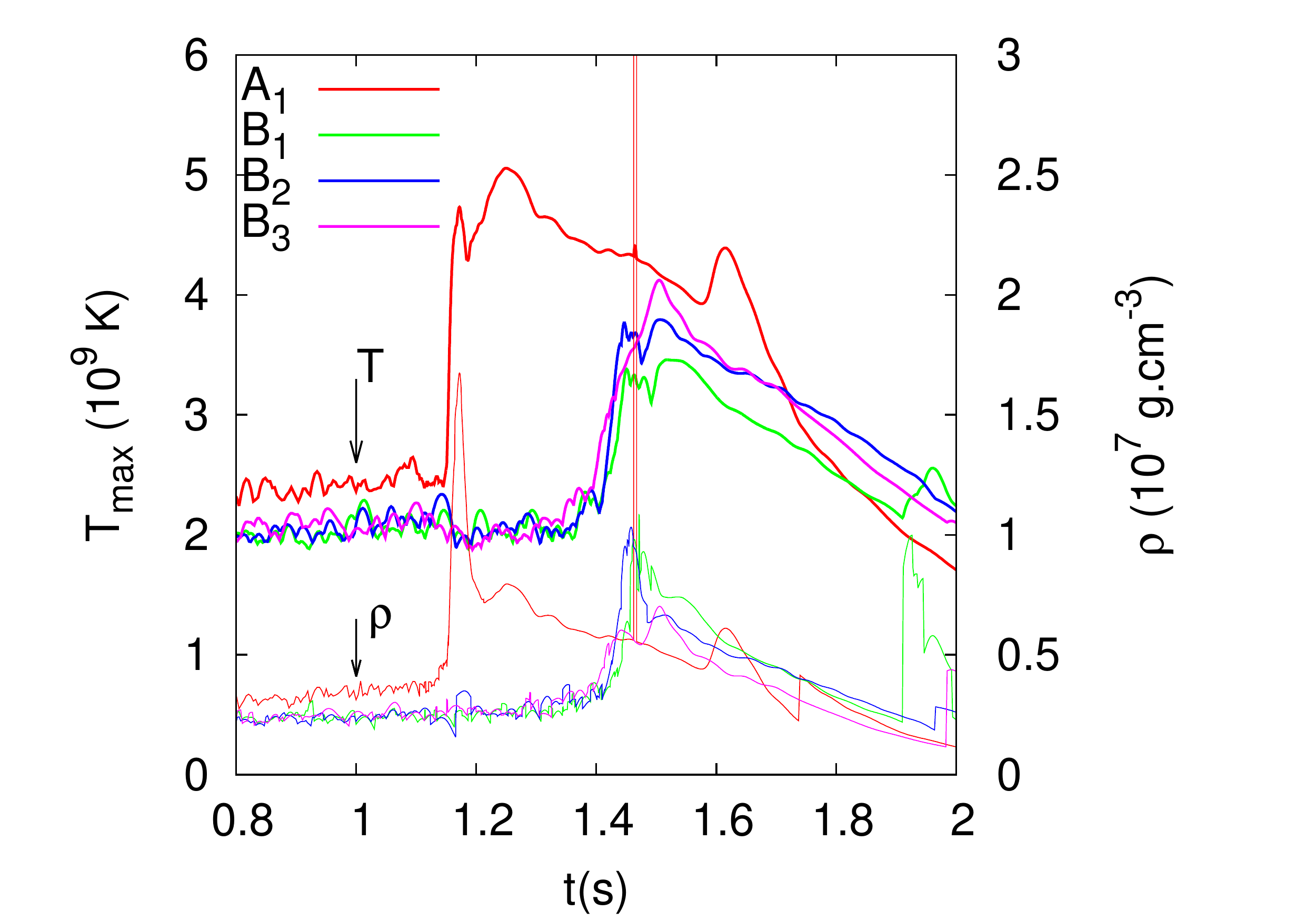}
\caption{Maximum temperature $T_{max}$ and its corresponding density $\rho(T_{max})$ for models A$_1$, B$_1$, B$_2$~and B$_3$. We show here the values achieved by any SPH-particle with CO composition as a function of the elapsed time.}  
\label{fig4}
\end{figure}

\begin{figure}
l\includegraphics[angle=0,width=\columnwidth]{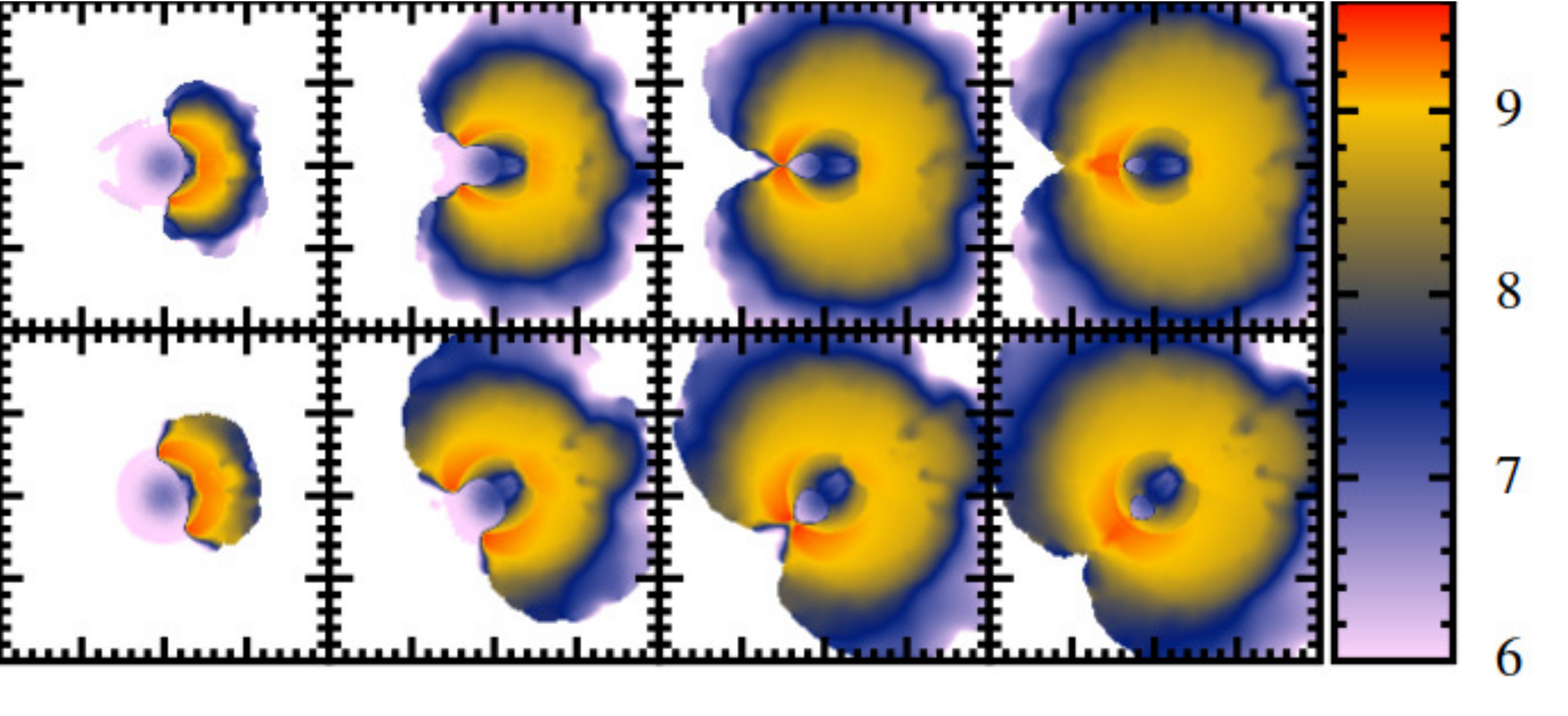}
\caption{Temperature colormap in a (comoving) XY-slice (upper row) and a (static) YZ-slice (lower row), showing the explosion of the helium envelope of model $B_3$ in Table~\ref{table1} at times $t=0.707$, $1.173$, $1.399$, and $1.513$ s. The rotation of the WD is well noticeable in the YZ slices. We note how the collision of the detonation waves at the antipodes takes place at quite different times in both slice sequences. The box size is $[-2:2]~10^4$~km in all directions.}
\label{fig5}
\end{figure}

\begin{figure}
\includegraphics[angle=0,width=\columnwidth]{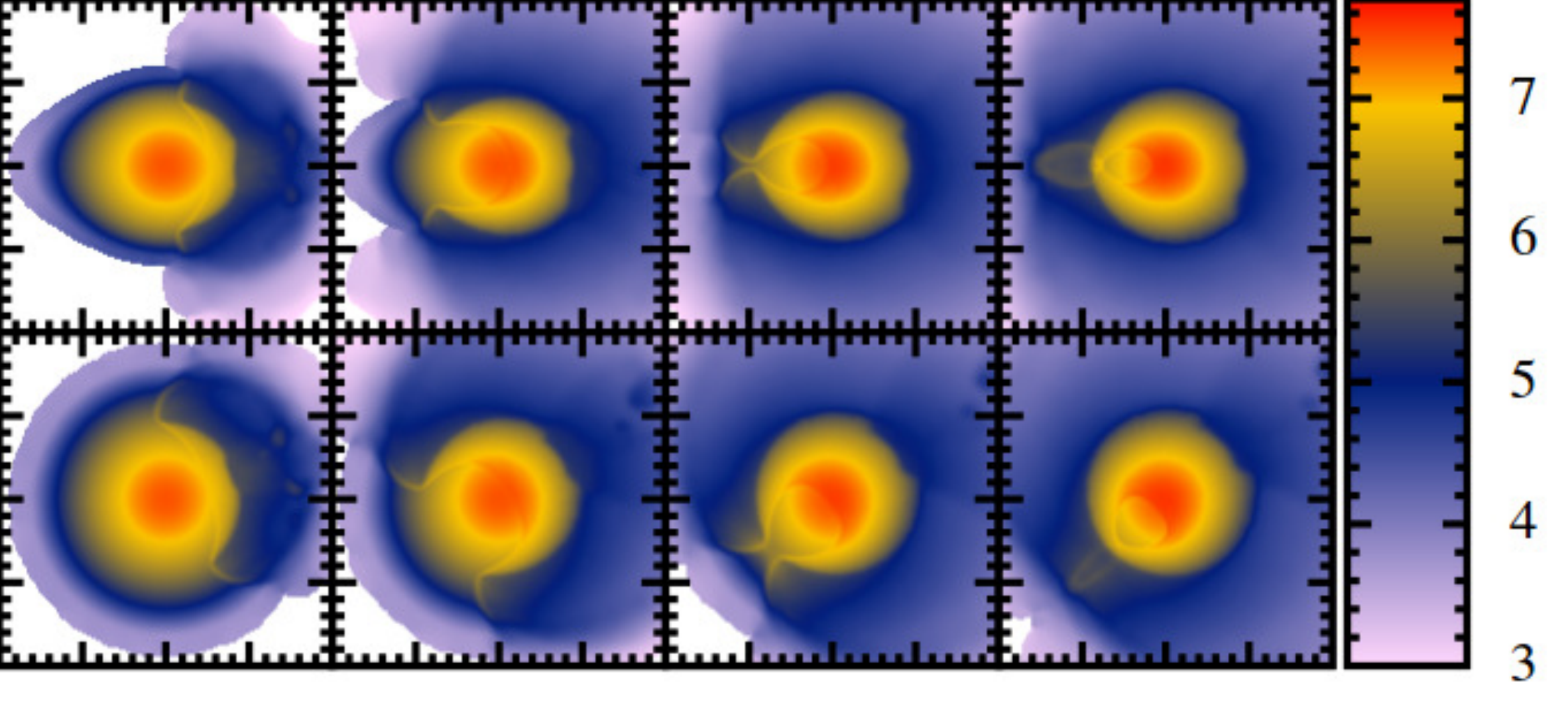}
\caption{Same as Figure \ref{fig5} but for density. The box size is $[-1:1]~10^4$~km in all directions.}
\label{fig6}
\end{figure}

\begin{figure}
\includegraphics[angle=0,width=\columnwidth]{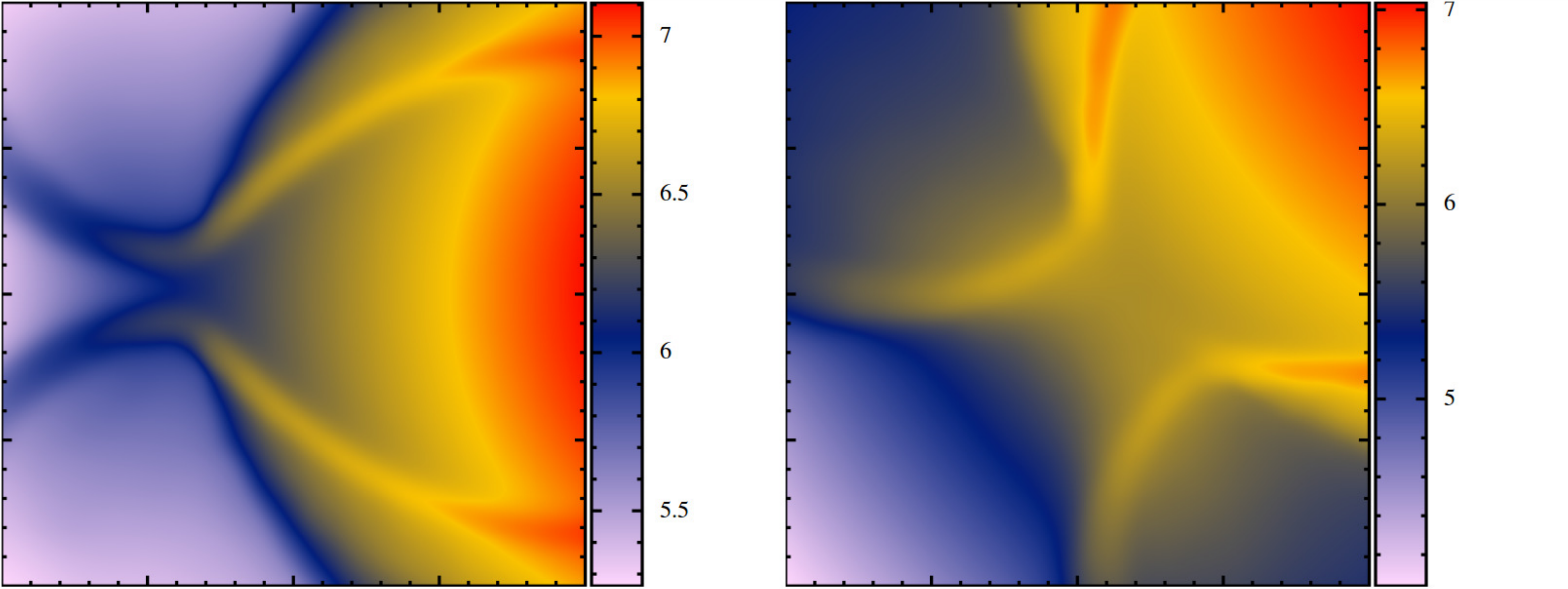}
\caption{Colormap of density around the convergence region at time $t=1.399$~s showing the time-shifting among wave arrivals in the polar plane XY (left) and equatorial YZ (right).}
\label{fig7}
\end{figure}

\begin{figure}
\includegraphics[angle=0,width=1\columnwidth]{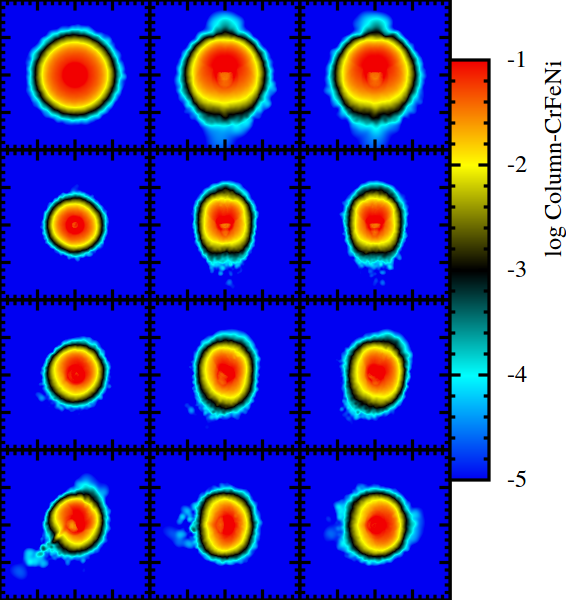}
\caption{From left to right: column densities of the radioactive  $^{48}\mathrm{Cr}$+$^{52}\mathrm{Fe}$+ $^{56}\mathrm{Ni}$~mass fractions along the X (polar view), Y, and Z directions at times $t\simeq 8.57, 8.07, 8.32, 8.12$~s for models A$_1$, B$_1$, B$_2$, and B$_3$ (from top to bottom), respectively. The boxes have a side length of $4~10^{5}$~km }  
\label{fig8}
\end{figure}

In order to produce an amount of $^{56}\mathrm{Ni}$ compatible to what is observed in a standard SNe Ia explosion it is also necessary to get the detonation of the CO core. According to our results (see  Fig.~\ref{fig4}), the core detonation is also the most probable outcome, even  when the WD rotates fast, close to the centrifugal breaking.      

\begin{deluxetable*}{ccccrccccc}
\tablecaption{Yields synthesized during the combustion of the He-shell (in M$\sun$)}
\tablehead{
\colhead{\quad}&\colhead{A$_1$}&\colhead{B$_1$}&\colhead{B$_2$}&\colhead{B$_3$}&\colhead{B$_4$}&\colhead{B$_5$}&\colhead{B$_6$} 
}
\startdata
$^4$He&$4.20~10^{-2}$&$5.47~10^{-2}$&$5.45~10^{-2}$ &$5.43~10^{-2}$&$6.31~10^{-2}$&$6.28~10^{-2}$&$6.23~10^{-2}$ \\
$^{12}$C&$1.92~10^{-4}$&$4.51~10^{-4}$&$4.34~10^{-4}$&$4.43~10^{-4}$&$2.99~10^{-4}$&$2.98~10^{-4}$&$3.01~10^{-4}$ \\
$^{16}$O&$5.35~10^{-7}$&$8.67~10^{-7}$&$8.57~10^{-7}$ &$8.57~10^{-7}$&$7.88~10^{-7}$&$7.80~10^{-7}$&$7.92~10^{-7}$ \\
$^{20}$Ne&$6.05~10^{-8}$&$1.24~10^{-7}$&$1.18~10^{-7}$ &$1.20~10^{-7}$&$9.22~10^{-8}$&$9.10~10^{-8}$&$9.15~10^{-8}$ \\
$^{24}$Mg&$4.77~10^{-7}$&$1.14~10^{-6}$&$1.07~10^{-6}$ &$1.09~10^{-6}$&$7.44~10^{-7}$&$7.34~10^{-7}$&$7.33~10^{-7}$ \\
$^{28}$Si&$5.02~10^{-6}$&$1.25~10^{-5}$&$1.22~10^{-5}$ &$1.23~10^{-5}$&$7.95~10^{-6}$&$7.97~10^{-6}$&$8.20~10^{-6}$ \\
$^{32}$S&$5.96~10^{-5}$&$1.26~10^{-4}$&$1.25~10^{-4}$ &$1.27~10^{-4}$&$9.26~10^{-5}$&$9.25~10^{-5}$&$9.80~10^{-5}$ \\
$^{36}$Ar&$9.74~10^{-4}$&$1.56~10^{-3}$&$1.55~10^{-3}$ &$1.52~10^{-3}$&$1.56~10^{-3}$&$1.55~10^{-3}$&$1.55~10^{-3}$ \\
$^{40}$Ca&$4.73~10^{-4}$&$7.22~10^{-4}$&$7.16~10^{-4}$ &$7.01~10^{-4}$&$7.72~10^{-4}$&$7.72~10^{-4}$&$7.62~10^{-4}$ \\
$^{44}$Ti&$2.24~10^{-2}$&$3.81~10^{-2}$&$3.79~10^{-2}$ &$3.75~10^{-2}$&$3.43~10^{-2}$&$3.41~10^{-2}$&$3.39~10^{-2}$ \\
$^{48}$Cr&$3.07~10^{-2}$&$1.70~10^{-2}$&$1.74~10^{-2}$ &$1.78~10^{-2}$&$4.15~10^{-2}$&$4.16~10^{-2}$&$4.15~10^{-2}$ \\
$^{52}$Fe&$8.67~10^{-3}$&$1.06~10^{-3}$&$1.12~10^{-3}$ &$1.28~10^{-3}$&$1.05~10^{-2}$&$1.07~10^{-2}$&$1.12~10^{-2}$ \\
$^{56}$Ni&$1.25~10^{-3}$&$1.18~10^{-4}$&$1.48~10^{-4}$ &$1.90~10^{-4}$&$1.25~10^{-3}$&$1.35~10^{-3}$&$1.63~10^{-3}$\\
$^{60}$Sn&$1.83~10^{-5}$&$1.39~10^{-6}$&$1.63~10^{-6}$ &$2.42~10^{-6}$&$1.94~10^{-5}$&$2.11~10^{-5}$&$2.60~10^{-5}$
\enddata
\label{table3}
\end{deluxetable*}

\subsection{He-shell detonation: Increasing the ignition density at the core-envelope interface.}
The precise value at which the first sparks of helium ignite has a strong impact on some of the yields coming from the detonation of the helium shell. The reference models B$_1$, B$_2$, and B$_3$ discussed above assumed a low ignition density value, $\rho_{He}=1.1~10^6$~\dens, close to the minimum necessary to build a steady detonation. The impact of raising the ignition density of helium at the interface up to $\rho_{He}=1.5~10^6$~\dens~ is explored in models B$_4$, B$_5$, and B$_6$. As the base of the He-shell \ is moved deeper its mass and thickness increases so that the total mass of the WD remains constant (see Table~\ref{table1}). The combination of a higher ignition density and a more massive envelope, (i.e a larger explosion tamper) leads to higher combustion temperatures, thus favoring the synthesis of iron group elements. In particular, the $^{52}\mathrm{Fe}$~and $^{56}\mathrm{Ni}$ yields are increased by a factor of ten (Table~\ref{table2}) while the released nuclear energy rises a $50\%$ (Table~\ref{table4}). The largest amount of Fe-Ni is synthesized in the off-axis igniter B$_6$, whereas the aligned igniter, model B$_4$, gives an amount of IGE similar to those of the non-rotating model A$_1$. 
  
\begin{figure}\includegraphics[angle=-90,width=\columnwidth]{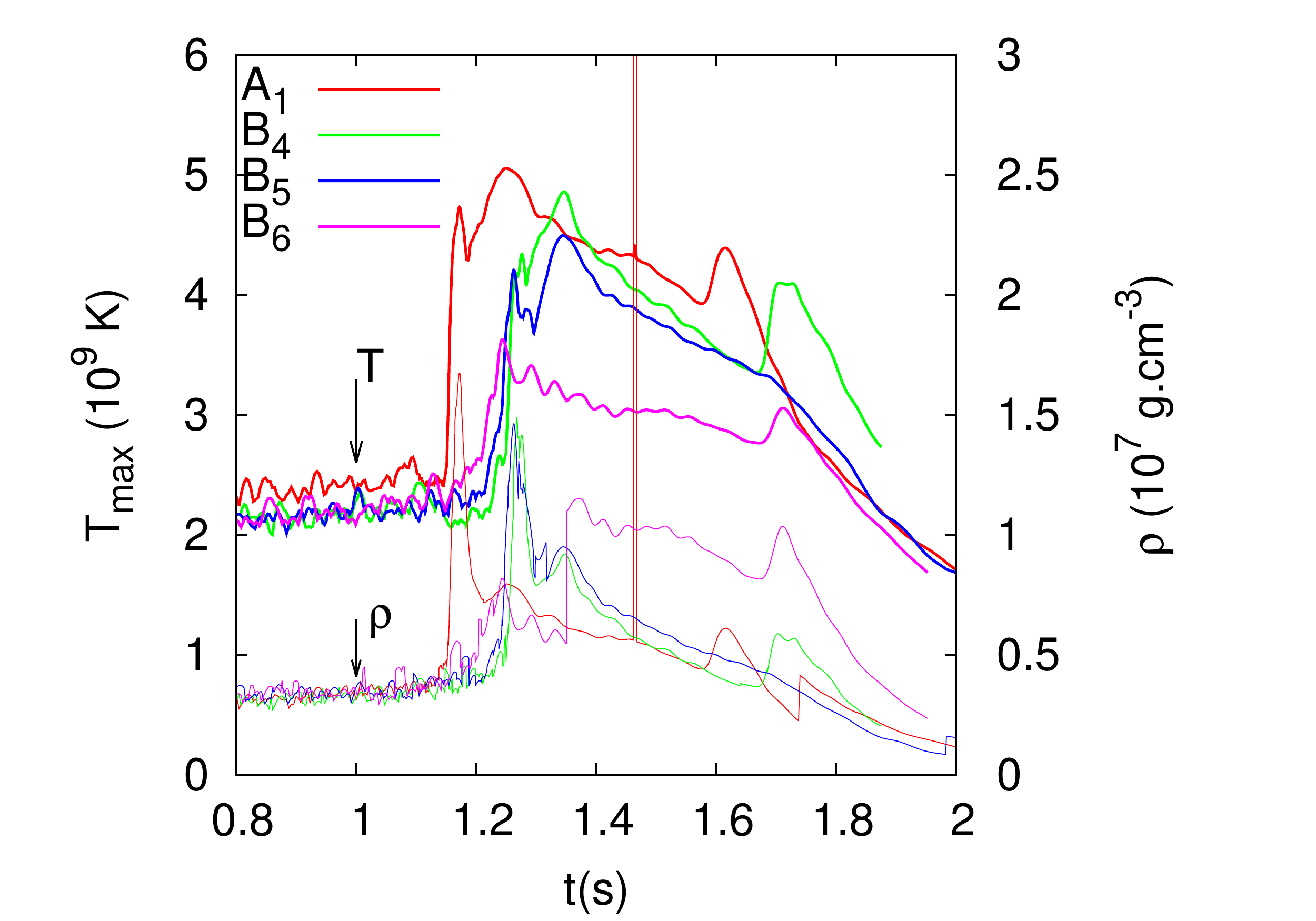}
\caption{Maximum temperature $T_{max}$ and its corresponding density $\rho(T_{max})$ for models A$_1$, B$_4$, B$_5$~and B$_6$. As in Fig.~\ref{fig4}, we show here the values achieved by any SPH-particle with CO composition as a function of the elapsed time.}  
\label{fig9}
\end{figure}

The evolution of $T_{max}$ (maximum temperature in the core with the nuclear reactions turned-off) and $\rho(T_{max})$ of models B$_4$, B$_5$, and B$_6$ is shown in Fig.~\ref{fig9}. The maximum temperature and densities achieved in models B$_4$ and B$_5$ are larger than in models B$_1$ and B$_2$. Even though model B$_6$ has a peak of $T_{max}$ similar to that of B$_3$ the evolution of $\rho (T_{max})$ is quite different because it has an extended plateau where $\rho (T_{max})\simeq 10^7$~\dens~ between $1.4$ and $1.7$~s. Therefore, the conditions to induce the detonation of the core are even more favorable in models B$_4$, B$_5$, and B$_6$ than in models B$_1$, B$_2$, and B$_3$, that ignite helium at a lower density at the interface.     
 
\subsection{Models with a thinner He-layer}

One historical objection to the subCh-mass route to SNe Ia is that it predicts a too large nickel production in the high-velocity external layers, which is not seen in the spectra. As suggested by \cite{bil07}, one remedy is to consider thinner helium envelopes so that the amount of synthesized $^{56}\mathrm{Ni}$ is proportionally reduced. But this poses a problem to the robustness of the DDet mechanism, as it may not work below some critical mass of the envelope. Nevertheless, several multi-D studies have shown that the DDet mechanism may work even for envelopes as low as $\Delta He \simeq 0.01 M_{\sun}$~\citep{fin10,sim12}. It is worth noting that SNe Ia may also arise from the violent merger of two massive CO-WDs capped with tiny helium shells, $\simeq 0.005$~M$_\sun$ each \citep{gui10,pak13}. Hydrodynamic simulations by \cite{pak13} predict that the He-detonation may induce the detonation of the, assumed non-rotating, CO core.  Thus, the explosion mechanism invoked in this double degenerate model is rather similar to the DDet mechanism on rotating WDs presented in this work.           

We have studied three additional cases, namely C$_1$, D$_1$ and D$_2$ in Table~\ref{table1}, with $\Delta M_{He} \simeq 0.05~M_{\sun}$, which is half of the He-envelope mass used in B-models. Model C$_1$ is the new non-rotating control case with spherical symmetry, central density $\rho_c = 6.82~10^7$\dens, and $\Delta M_{He}=0.052~M_{\sun}$. The rotating models, D$_1$ and D$_2$, spin with $w=0.65$~s$^{-1}$~and have a central density $\rho_c = 6.87~10^7$\dens~ and $\Delta M_{He}= 0.053 M_{\sun}$. In spite of having a larger rotational velocity, D-models are not as oblated as B-models because they are more massive (see Table~\ref{table1}). 

The evolution of cases C$_1$, D$_1$ and D$_2$ is similar to that of models with thicker helium envelopes. Table \ref{table2} presents a summary of the results. Again, the maximum temperature $T_{max}$~and $\rho(T_{max})$~(estimated with the $^{12}$C$+^{12}$C~reaction turned off) achieved by a carbon particle at the antipodes, is high enough to induce the detonation of the core. If the density at the edge of the core is similar for all models, then the energy released during the evaporation of the helium envelope roughly scales with the mass of the He-shell (Table \ref{table2}) 

\section{Core detonation}
\label{coredet}

Now we compute models A$_1$, B$_4$, B$_5$ and B$_6$ in Table~\ref{table1}, allowing the binary $^{12}C+^{12}C$ and $^{16}O+^{16}O$ reactions to proceed. In all cases the spontaneous detonation of the core and the complete destruction of the WD is   obtained. The released nuclear energy, final kinetic energy and the rough nucleosynthesis do match the most basic SNe Ia observational constraints. A summary of these magnitudes is provided in Table~\ref{table4}. 
 
\begin{deluxetable}{ccccrccccc}
\tablecaption{Main features during the complete detonation of the WD.}
\tablehead{
\colhead{Model}&\colhead{$E_{kin}$}&\colhead{$E_{nuc}$}&\colhead {$IME$}&\colhead{$IGE$}& \colhead{$^{56}\mathrm{Ni}$} \cr
  &\colhead{$10^{51}$~ergs}&\colhead{$10^{51}$~ergs}&\colhead{$M_{\sun}$}&\colhead{M$_{\sun}$}&\colhead{M$_{\sun}$} \\
}
\startdata
A$_1$&$ 1.09$&$ 1.22$&$0.36$&$0.45$&$0.37$ \\
B$_4$&$1.24$&$1.40$&$0.38$&$0.52$&$0.42$ \\
B$_5$&$1.27$&$1.43$ &$0.38$&$0.54$&$0.44$ \\
B$_6$&$1.26$&$1.41$&$0.38$&$0.53$&$0.43$ \\
\enddata
\tablecomments{Main features of the complete detonation of models A, B$_4$, B$_5$ and B$_6$~at $t=11.5$~s.   
}
\label{table4}
\end{deluxetable}

The complete explosion of the spherically symmetric model A$_1$~is in agreement  with the evolution of similar models calculated by other groups. For example, the obtained Ni yield, $0.38$~M$_\sun$~is almost equal to that obtained by \cite{mol13} for a similar model (their model A). The kinetic energy at $t\simeq 11.5$~s is $\simeq 1.1~10^{51}$~ergs, completely compatible with a standard SNe Ia explosion.  

\begin{figure}
\includegraphics[angle=0,width=\columnwidth]{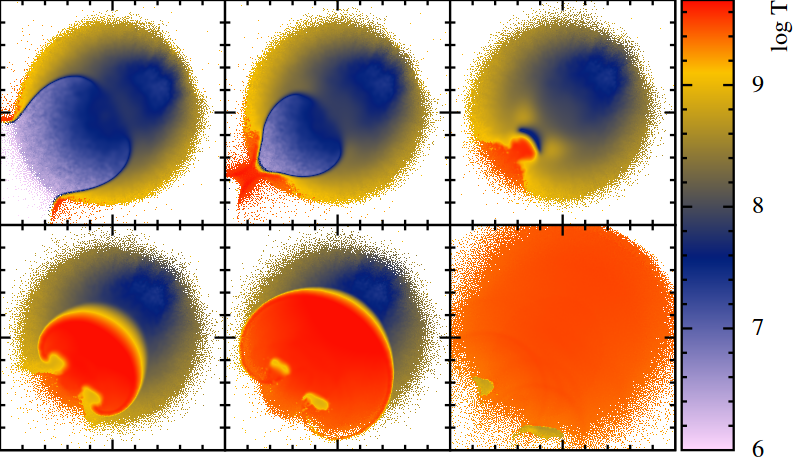} 
\caption{Colormap of temperature in a YZ (equatorial) slice, showing the core detonation of model B$_6$ in Table~\ref{table1} at times $t=1.10, 1.22, 1.42, 1.62,  1.72$, and $2.03$~s, respectively. The box size is $[-5:5]~10^3$~km in all directions. }
\label{fig10}
\end{figure}

\begin{figure}
\includegraphics[angle=0,width=\columnwidth]{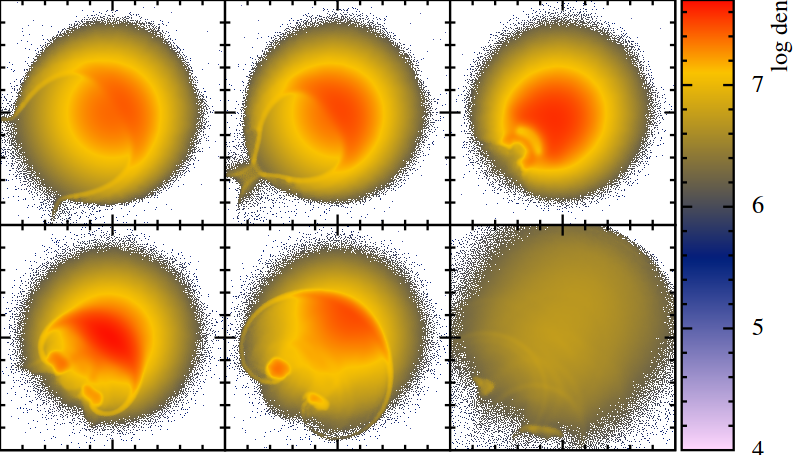}
\caption{Same as Fig.~\ref{fig10}, but for density. 
}  
\label{fig11}
\end{figure}

Several snapshots showing the detonation of the core of model B$_6$ are depicted in the equatorial slice shown in Figures~\ref{fig10} and \ref{fig11}. A hot-spot appears at the antipodes, when the He-shell ashes converge at $t\simeq 1.22$~s (second snapshot). Nevertheless, the spontaneous detonation of the core still has to wait until $t\simeq 1.42$~s, moment at which the compression waves arriving from the hot-spot and from the center of the WD meet (see the third snapshot in figure \ref{fig11}). After this moment a steady detonation forms and propagates inwards through the core (first and second snapshots in the second row). In the meanwhile, the rotation of the core between the first and fifth snapshots is clearly visible. Finally, the whole core has been burnt at the last snapshot at $t=2.03$~s. The detonation of the core in models B$_4$ and B$_5$ follow a qualitatively similar path.

\begin{figure}
\includegraphics[angle=-90,width=\columnwidth]{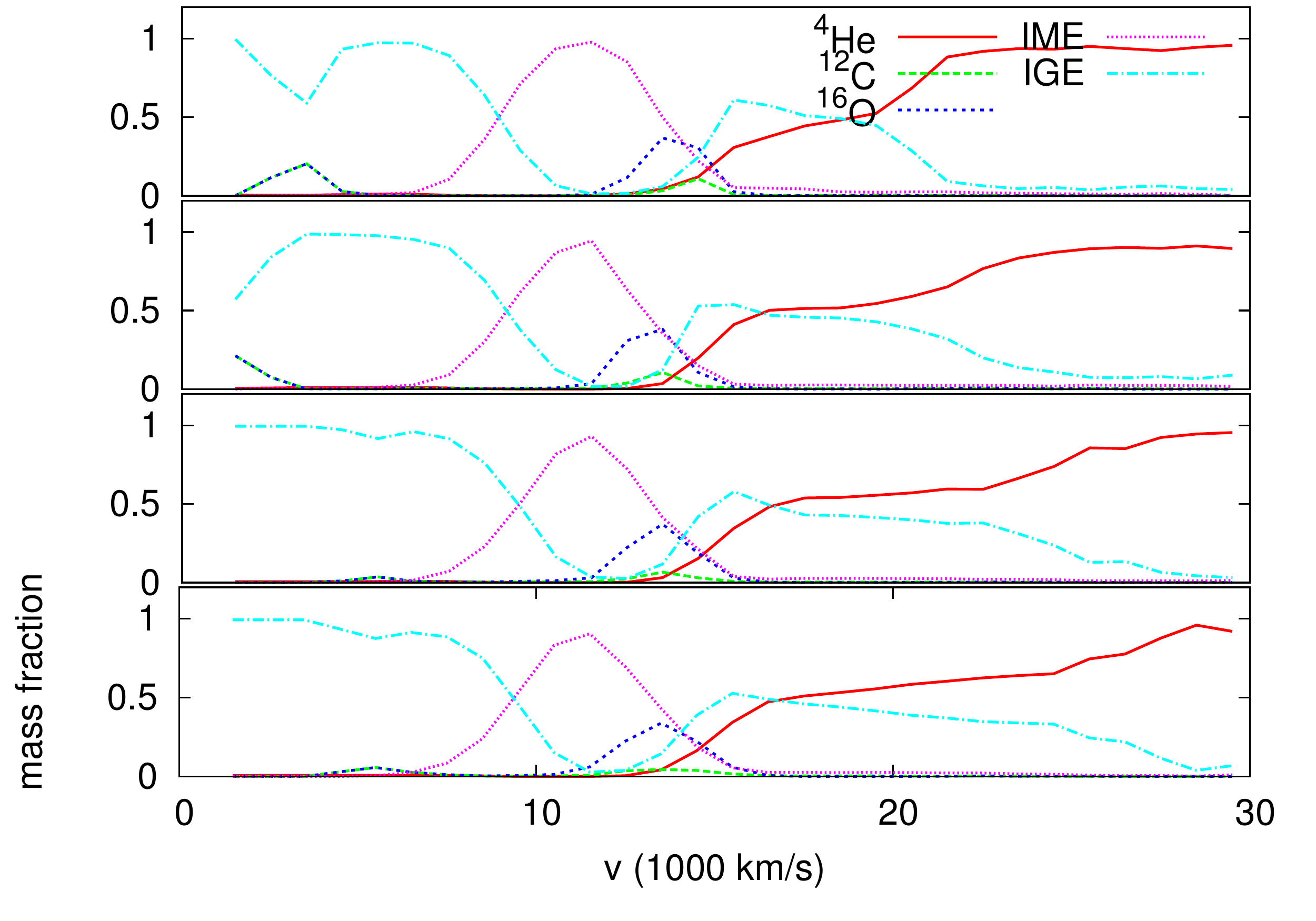}
\caption{Complete detonation of the WD: mass fractions of the main groups of nuclei in velocity space at $t=11.5$~s. From top to bottom: models A$_1$, B$_4$, B$_5$, and B$_6$ (see  Tables~\ref{table1} and \ref{table5}).  }  
\label{fig12}
\end{figure}

Because the CO core is incinerated supersonically, we do not expect large differences in the energetics or in the ejected nuclear yields among the rotating models and, in fact, this is what our simulations show (Table \ref{table4}). There are, however, several differences in the product yields of the explosion with respect those of the spherically symmetric model A$_1$. While the amount of intermediate-mass elements (IME) is only slightly larger in the rotating models, which can be explained by the larger mass of the WD, the iron-group elements\footnote{We have grouped all nuclei between $^{20}\mathrm{Ne}$~and $^{40}\mathrm{Ca}$~as IME and from $^{44}\mathrm{Ti}$~ up to $^{60}\mathrm{Zn}$~as belonging to the IGE.} (IGE) are comparatively more copiously produced. This is a different trend as that found in rotating Chandrasekhar and Super Chandrasekhar-mass models igniting at much higher densities. In those models, a fast rotation favors the production of the IME elements \citep{pfa10a,pfa10b}. The enhanced production of IGE in models B$_4$, B$_5$, and B$_6$~with respect the non-rotating model $A_1$~is due to  their larger core and He-shell masses. Having a thicker He-envelope tamper increases the average density of the core at the moment of Carbon-detonation which ultimately favors the production of heavy nuclei. Because of the larger production of IGE, the kinetic energy of the explosion is consequently larger in the rotating models, $\simeq 1.25~10^{51}$~ergs.  The distribution of the abundances in velocity space at $t=11.5$~s is depicted in Figure \ref{fig12}. The most relevant feature is that the IGE profiles (dashed-blue lines) spread to  larger velocities in the rotating models, especially in the oblique ignitors B$_5$ and B$_6$. 

\begin{deluxetable}{ccccrccccc}
\tablecaption{Yields synthesized during the complete detonation of the WD (in M$\sun$)  }
\tablehead{
\colhead{\quad}&\colhead{A$_1$}&\colhead{B$_4$}&\colhead{B$_5$}&\colhead{B$_6$}&  
}
\startdata
$^4$He&$4.35~10^{-2}$&$6.63~10^{-2}$&$6.62~10^{-2}$ &$6.58~10^{-2}$ \\
$^{12}$C&$2.71~10^{-2}$&$2.25~10^{-2}$&$1.63~10^{-2}$&$2.07~10^{-2}$ \\
$^{16}$O&$8.74~10^{-2}$&$8.89~10^{-2}$&$7.85~10^{-2}$ &$8.45~10^{-2}$ \\
$^{20}$Ne&$2.64~10^{-3}$&$2.74~10^{-3}$&$2.30~10^{-3}$ &$2.38~10^{-3}$ \\
$^{24}$Mg&$2.58~10^{-2}$&$2.83~10^{-2}$&$2.66~10^{-2}$ &$2.73~10^{-2}$ \\
$^{28}$Si&$1.56~10^{-1}$&$1.72~10^{-1}$&$1.69~10^{-1}$ &$1.69~10^{-1}$ \\
$^{32}$S&$9.55~10^{-2}$&$1.03~10^{-1}$&$1.03~10^{-1}$ &$1.10~10^{-1}$ \\
$^{36}$Ar&$3.03~10^{-2}$&$3.23~10^{-2}$&$3.28~10^{-2}$ &$3.19~10^{-2}$ \\
$^{40}$Ca&$4.39~10^{-2}$&$4.58~10^{-2}$&$4.62~10^{-2}$ &$4.47~10^{-2}$ \\
$^{44}$Ti&$2.47~10^{-2}$&$3.46~10^{-2}$&$3.45~10^{-2}$ &$3.42~10^{-2}$ \\
$^{48}$Cr&$3.38~10^{-2}$&$4.43~10^{-2}$&$4.37~10^{-2}$ &$4.35~10^{-2}$ \\
$^{52}$Fe&$2.16~10^{-2}$&$2.46~10^{-2}$&$2.57~10^{-2}$ &$2.56~10^{-2}$ \\
$^{56}$Ni&$3.66~10^{-1}$&$4.16~10^{-1}$&$4.36~10^{-1}$ &$4.30~10^{-1}$ \\
$^{60}$Sn&$7.72~10^{-4}$&$1.12~10^{-3}$&$9.29~10^{-4}$ &$9.08~10^{-4}$ \\
\enddata
\label{table5}
\end{deluxetable}

\section{Conclusions}

In this work, we addressed the question of the fate of rotating white dwarfs that detonate helium at the base of an accreted shell, when their masses are well below the Chandrasekhar-mass limit. A study of this kind has never be attempted before, being pertinent by several reasons. The more compelling of them being that in a spinning WD the location of the initial kernel/s leading to the Helium-shell detonation are not necessarily located on the rotation axis. Thus, the strong (almost point-like) convergence of the ashes of the He-detonation on the antipodes of the igniting region, typical of the spherically symmetric models, is lost. Such loss of focusing in the convergence of the ashes changes the physical conditions at the underlying carbon core, which may be less prone to detonate. A second goal was to make a comparison among the main observables coming from both, the rotating and non-rotating models. To do that we have considered two potential explosion scenarios. In the first case the secondary carbon-detonation was artificially suppressed and the main observables of the sub-luminous event, produced by the He-shell detonation, were determined. In the second case carbon was allowed to detonate which, according to our own results is the most plausible outcome. Again, the main observables were obtained and compared with a non-rotating spherically symmetric model. 

The rotational velocity of an accreting WD is set by the total amount of accreted material, by the efficiency of angular momentum transport from the surface to the core, and by the angular momentum losses. In the case of the DDet scenario the mass of the He-shell is not as large as in the Chandrasekhar-mass models of Type Ia supernova and the ensuing angular velocity is expected to be lower. The precise profile of the angular velocity in the progenitor of subCh-mass explosion models is not well known (see Sect. \ref{SecRotation}). On a practical basis, we have adopted rigid rotation which facilitates building rotating equilibrium models with the SPH technique,  being a realistic hypothesis in case of efficient angular momentum transport. In any case, our simulations aim to study how the propagation of the helium-detonation is affected by a change in the geometry of the He-shell and the CO-core interface. Assuming rigid rotation is enough to conduct such exploratory study.

As a principal result, we confirm the robustness of the DDet mechanism as a viable scenario to give rise a SNe Ia explosion. According to our results, igniting helium far from the rotational axis blurs the convergence of the detonation to the antipodes, as expected. But, rather than hindering it, the slight asynchronicity in the arrival of the detonation waves seems to enhance the chances of inducing the carbon detonation below the CO core (see Fig.~\ref{fig4}). When the helium initially detonates close to the rotational axis the geometrical focusing at the antipodes is preserved and the results are similar to those of the spherically symmetric model. These results also hold for smaller helium shells, $\simeq 0.05$~M$_\sun$ (D-models in Table~\ref{table1}).

We have carried out a separate study of both, the detonation of the He-shell alone and the combined He-shell and CO-core detonations. The former case would give rise to a peculiar sub-luminous SNe Ia event, in which the light curve is powered by the radioactive $^{48}\mathrm{Cr}$~and $^{52}\mathrm{Fe}$, with a minor contribution of $^{56}\mathrm{Ni}$. Nevertheless, we found that the precise yield of $^{56}\mathrm{Ni}$~is very dependent on the density at the base of the He-shell at the moment of the explosion. The radioactive $^{44}\mathrm{Ti}$~seems to be more copiously produced in rotating WDs. The column-density map of the radioactive elements produced in the explosion of the spinning models shows a larger loss of the spherical symmetry than in the non-rotating case (see Fig.~\ref{fig8}). Such asymmetry might increase the polarization signatures of the spectra, which is low in standard non-rotating subCh-mass models \citep{bul16}. Nonetheless, this qualitative result has to be confirmed with more detailed calculations of the polarization spectra. 

When the $^{12}$C$+^{12}$C reaction is allowed to proceed, the detonation of the He-shell is always followed by the spontaneous detonation of the core. A robust explosion, energetically compatible with a standard SNe Ia event, is obtained in all the cases studied. However, the rotating models do show an enhanced production of IGE, some of them moving at a large velocity during the homologous expansion. The larger amount of $^{56}\mathrm{Ni}$~moving at $v\ge 2~10^4$~km~s$^{-1}$, besides the expected increase in the polarization signatures of the explosion, conspire against fast spinning WDs with thick helium layers as a viable progenitors of SNe Ia. A reduction in the mass of the accreted He-shell would help with this problem. We have shown that halving the mass of the helium envelope still leads to the detonation of the core in spinning WDs (D-models in Table~\ref{table1}). Nevertheless, reducing the mass of the accreted envelope also lowers the amount of angular momentum gained by the WD. For He-shell masses $\le 0.01$~M$_\sun$~the geometry of the WD would remain almost spherical. 

The combination of a low-mass He-shell on top of an oblated substrate made of carbon and oxygen may, however, be realized in the Double-Degenerate scenario \citep{gui10,dan15}. It has been suggested that the DDet mechanism, postulated to explain the subCh-mass route to SNe Ia, could also be at work in the DD scenario \citep{pak13}. In this regard, the results presented in this manuscript can also be useful to better understand the double degenerate scenario of Type Ia Supernova.       

\acknowledgements

We acknowledge useful comments by Stan Woosley. This work has been supported by the MINECO Spanish project AYA2017-86274-P (DG), by the Swiss Platform for Advanced Scientific Computing (PASC) project SPH-EXA (RC and DG) and by the 
 Spanish MINECO-FEDER project AYA2015-63588-P (ID). The authors acknowledge the support of sciCORE (http://scicore.unibas.ch/) scientific computing core facility at University of Basel, where some of the calculations were performed.

\clearpage
\appendix
\section{Implementation of rotation.}

An accurate method to build rotating WDs in hydrostatic equilibrium within the SPH framework does not exist. We have developed and checked a relaxation procedure which is able to produce self-gravitational rotating structures in equilibrium. We assume that rotation is axisymmetric and that any physical and geometrical feature of the oblated structure in equilibrium is basically determined by the total mass $M\textsubscript{WD}$ and total angular momentum $J\textsubscript{WD}$. Both magnitudes, $M\textsubscript{WD}$~and $J\textsubscript{WD}$, are specified at $t=0$ and kept constant during the relaxation process, during which we let the sample of SPH mass points evolve under the self-gravity and the centripetal force in a co-rotational frame. After several sound-crossing times the rotating structures come to an equilibrium.

\begin{figure}
\includegraphics[angle=-90,width=\columnwidth]{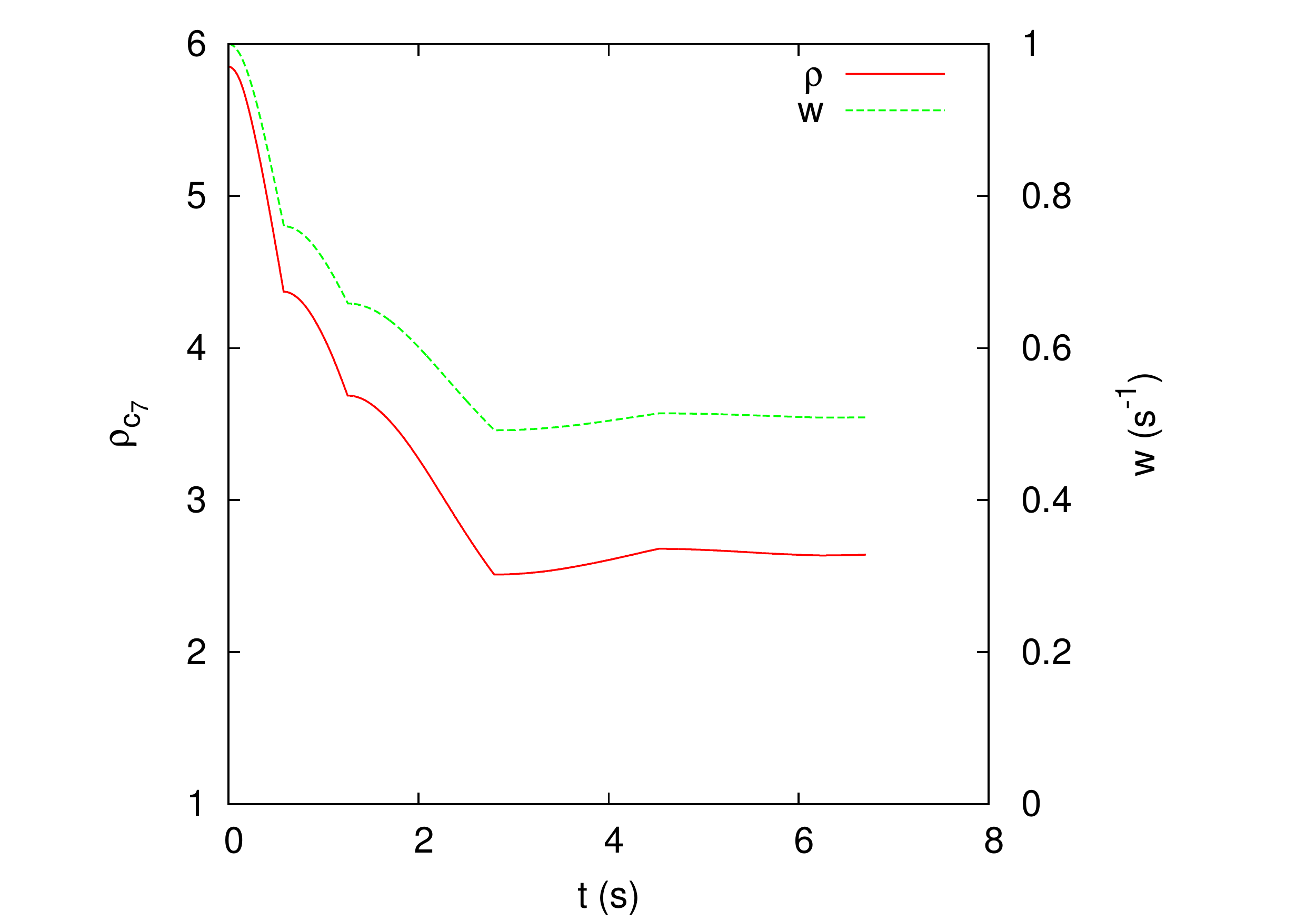}
\caption{Example of a relaxation sequence towards equilibrium. The total mass $M=1.081$~M$_\sun$~and total angular momentum $J=0.798~10^{50}$~erg.s were kept constant, while the angular velocity and central density evolve to achieve stable values. The locations where the slope of the curves change indicate the times at which the particle velocities are set to zero to remove numerical noise.}  
\label{fig13}
\end{figure}

Starting from a spherically symmetric model of a WD with central density $\rho_0$, a sample of $N$ mass-particles is spread according to the density profile $\rho(r)$. The distribution in the spherical angles $\phi$ and $\theta$ is chosen at random. We introduce the rigid rotation as a fictitious centripetal force, which is added to the gravity ${\bf f_c} = -{\bf\omega}(t)\times({\bf\omega}(t)\times {\bf r}(t))$, where ${\bf \omega}(t) = \omega_x (t)\hat{\mathrm{\bf{i}}}$ is the angular velocity at the elapsed time $t$ (the $X$-axis has been assumed as the rotation axis in this work) and ${\bf r}(t)$, the position vector of the particle. The angular momentum of this configuration, as view from an inertial reference frame, is $J_x=I_{xx}~\omega_x$ where $I_{xx}$ is the moment of inertia around the rotation axis. We let this configuration free to evolve and compute the time-dependent angular velocity of the WD at each integration step, so that the total angular momentum is preserved $\omega_x(t)=\frac{I_{xx}(t=0)}{I_{xx}(t)} \omega_x(t=0)$. The velocity of the particles is regularly set to zero to remove the spurious numerical noise. A typical relaxation sequence is shown in Figure \ref{fig13}, where we see how the $\omega(t)$ and the central density $\rho(t)$ approach stable values after several seconds of evolution. A summary of the equilibrium rotation features of the WDs used in this work is provided in Table~\ref{table6}.              

\begin{deluxetable}{ccccrccccc}
\tablecaption{Rotation features of the  white dwarfs after relaxation.}
\tablehead{
\colhead{Model}&\colhead{Mass}&\colhead{J}&\colhead {$\omega$}&\colhead{W}& \colhead{T}&\colhead{U}&\colhead{T/$\vert$W$\vert$}&\colhead{$R_e$}&\colhead{${R_p}/R_e$} \cr
  &\colhead{$M_{\sun}$}&\colhead{$10^{50}$~erg.s}&\colhead{$s^{-1}$}&\colhead{$10^{50}$~erg}&\colhead{$10^{50}$~erg}&\colhead{$10^{50}$~erg}& \colhead{-} &\colhead{km} &\colhead{-} \\
}
\startdata
B$$&$ 1.081$&$ 0.798$&$0.500$&$-4.805$&$0.200$&$2.710$& $0.042$&$8000$&$0.650$ \\
D$$&$ 1.187$&$ 0.619$&$0.650$&$-7.720$&$0.201$&$4.870$& $0.026$&$5560$&$0.797$ \\
\enddata
\tablecomments{Main features of the rigid rotators B and D of Table \ref{table1} at equilibrium. Symbols J, $\omega$~are the total angular momentum and the angular velocity whereas W, T, and U are the total gravitational, kinetic and internal energies. The two last columns show the equatorial radius $R_e$~and the polar to equatorial radii ratio, respectively. 
}
\label{table6}
\end{deluxetable}
\bibliographystyle{apj}
\bibliography{bibliography_r}


\end{document}